\input epsf.sty
\magnification=\magstep1
\looseness=2
\tolerance=1000
\def\ref{\par\noindent\hangindent 12pt}

\def\bsn{\medbreak\goodbreak\noindent}

\def\noi{\noindent}

\def\cl{\centerline}

\def\Mpc{$h^{-1}$~{\rm Mpc}}
\def\dlt#1{{#1}}
\def\je#1{{#1}}
\def\ed#1{{#1}}%editor's changes

%\font\big=cmr10 scaled \magstep1
%\font\smc=cmcsc10
%\font\small=cmr8
%\hsize=6.5 truein
%\vsize=9.5 truein
%\baselineskip=13pt plus1pt minus2pt
%\parskip=1pt plus1pt minus0pt
%\vglue 0.2truein
%\line{\hfill {\big Nature}, {\bf 385}, 139 - 141, 9 January 1997}
%\vskip 0.5cm

{\bf A 120--MPC PERIODICITY 
 IN THE THREE-DIMENSIONAL DISTRIBUTION
 OF GALAXY SUPERCLUSTERS} 

\bsn 
\cl{\bf J. Einasto$^1$, M. Einasto$^1$, S. Gottl\"ober$^2$, 
V. M\"uller$^2$, V. Saar$^1$, }
\cl{\bf A. A. Starobinsky$^3$, E. Tago$^1$, D. Tucker$^{2,4}$, 
H. Andernach$^{5,6}$, P. Frisch$^7$}

\bsn
\cl{$^1$Tartu Observatory, EE-2444 T\~oravere, Estonia}
\cl{$^2$Astrophysical Institute Potsdam, An der Sternwarte 16, D-14482
Potsdam, Germany}
\cl{$^3$Landau Inst. Theor. Physics, 117940 Moscow, Russia}
\cl{$^4$Fermilab, Batavia, IL 60510, USA}
\cl{$^5$INSA, ESA IUE Observatory, E-28080 Madrid, Spain}
\cl{$^6$Depto. de Astronom\'\i a, Univ. Guanajuato, Guanajuato, Mexico}
\cl{$^7$University Observatory, D-37083 G\"ottingen, Germany}

\bsn {\ed{\bf According to the favour models for the formation of large-scale
structure in the Universe (in which the dynamics of the Universe is dominated
by cold dark matter), the distribution of galaxies and clusters of galaxies
should be random on large scales. It therefore came as a surprise when a
periodicity was reported$^1$ in the distribution of high-density regions of
galaxies in the direction of Galactic poles, although the appearent lack of
periodicity in other directions led to the initial report being regarded as a
statistical anomaly$^2$. A subsequent study$^{3-6}$ also claimed evidence for
periodicity on the same scale, but the statistical significance of this result
was uncertain due to small number of clusters used.  Here, using a new
compilation$^7$ of available data on galaxy clusters, we present evidence for
a quasiregular three-dimensional network of rich superclusters and voids, with
the regions of high density separated by $\sim$ 120~Mpc. If this reflects the
distribution of all matter (luminous and dark), then there must exists some
hithero unknown process that produces regular structure on large scales.}}

During the past few years the number of clusters with measured redshifts has
increased considerably. To search for the possible presence of a regularity of
the distribution of matter in the Universe we have used a new compilation$^7$
of available data on rich clusters of galaxies catalogued by Abell and
collaborators{$^{8,9}$}.  The compilation has made use of all ($\sim$300)
published references on redshifts of both individual galaxies and Abell galaxy
clusters. Individual galaxies were associated \ed{with} a given Abell cluster
if they lay within a projected distance of $\leq$1.5~\Mpc\ (1 Abell radius)
and within a factor of two of the redshift estimated from the brightness of
the cluster's 10-th brightest galaxy, using the photometric estimate of
Peacock \& West$^{10}$ \je{($h$ is the Hubble constant in units of
100~km/s/Mpc)}.  The compilation contains measured redshifts for 869 of the
1304 clusters with an estimated redshift up to $z=0.12$. For the present
analysis we used all rich clusters (richness class $R\ge 0$) in this
compilation with at least two galaxy redshifts measured. The omission of the
435 clusters without measured redshifts does not affect our result because an
appropriate selection function was used.

\midinsert
\vskip 50mm
%{\epsfysize= 7 cm \epsfbox[75 450 275 700]{einf1.ps}}
{\epsfysize= 7 cm \epsfbox[75 230 275 480]{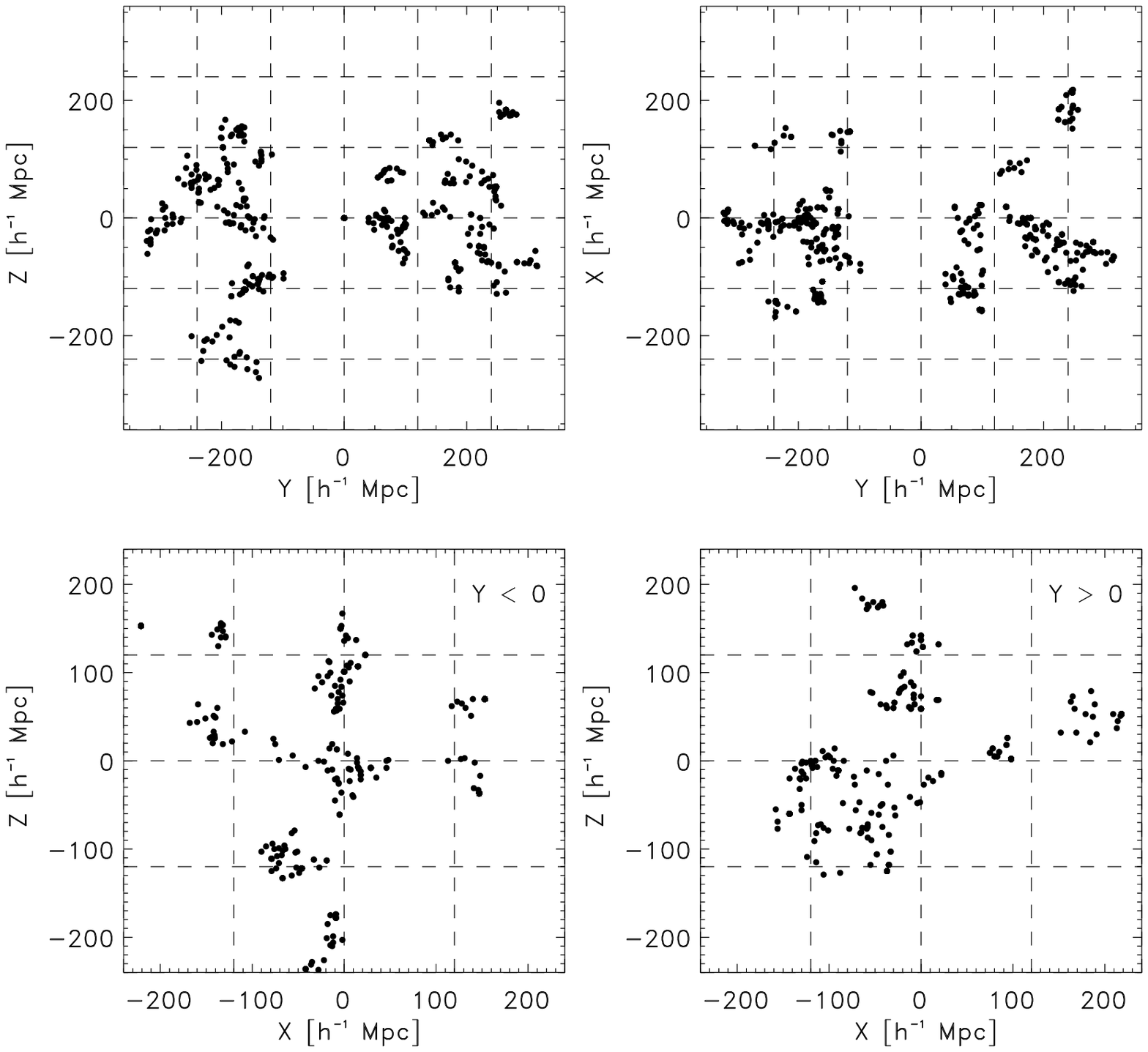}}
\bsn { FIG.~1.} The distribution of 319 clusters in 25 very rich
superclusters with at least 8 members (including 58 clusters with photometric
distance estimates), illustrating the network in the cluster distribution in
supergalactic coordinates. In the lower panels clusters in the northern and
southern Galactic hemispheres are plotted separately.  The supergalactic $Y=0$
plane coincides almost exactly with the Galactic equatorial plane, \ed{that
is,} with the zone of avoidance due to galactic absorption. The grid with step
size 120~\Mpc\ corresponds approximately to distances between high-density
regions across voids.  In the two upper panels and in the lower right panel
several superclusters overlap due to projection but are actually
well-separated in space.
\endinsert

This cluster sample (including clusters with estimated redshifts) was used to
construct a new catalogue of 220 superclusters of galaxies$^{11,12}$.  These
are systems of clusters where the distances between nearest neighbours among
member clusters do not exceed 24~\Mpc. In this way we find high-density
regions in the distribution of clusters of galaxies. In Fig.~1 we plot
clusters located in rich superclusters with at least 8 member clusters.  We
see a moderately regular network of superclusters and voids with a step size
of $\approx 120 \pm 20$~\Mpc\ where chains of superclusters are separated by
voids of almost equal size$^{11,12}$.  The whole distribution resembles a
three-dimensional chessboard$^{13}$.  Nearest neighbour test, and pencil-beam
and void analysis indicate that clusters in poor superclusters with less than
8 members and isolated clusters form a more uniform population, preferentially
located in void walls between rich superclusters but not filling the
voids$^{12}$.

To quantify the regularity of the cluster distribution we have
calculated the correlation function and the power spectrum for
clusters of galaxies.  The correlation function describes the
distribution of clusters in the real space, the power spectrum in the
\dlt{Fourier} space of density waves.  Analysis of various geometric 
models has shown that if superclusters form a quasiregular lattice
with an almost constant step size then the cluster correlation function is
oscillating, it has alternate secondary maxima and minima, separated
by half the period of oscillations. The period of spatial oscillations
of the correlation function is equal to the step size of the
distribution$^{14}$.  The amplitude of the power spectrum at the
wavelength corresponding to that period is enhanced with respect to
other wavelengths, that is, it is peaked. In contrast, if superclusters
are located randomly in space then the correlation function approaches
zero level at large separations and the power spectrum turns smoothly
from the region with positive spectral index on large wavelengths to a
negative index on small wavelengths$^{14}$.

\dlt{To calculate the power spectrum we have used the sample which 
contains clusters with measured redshifts only and lying in both
Galactic hemispheres out to the distance covered by our cluster and
supercluster catalogues.} The power spectrum was derived using two
different methods, a direct one where we calculate the distribution of
clusters in the wavenumber space, and an indirect method where we
first calculate the correlation function of clusters of galaxies and
then find the spectrum. In the latter case we make use of the fact
that the power spectrum and the correlation function are related by
the Fourier transform.

\midinsert
\vskip 20mm
%{\epsfysize= 7 cm \epsfbox[50 400 300 625]{einf2.ps}}
{\epsfysize= 7 cm \epsfbox[50 280 300 505]{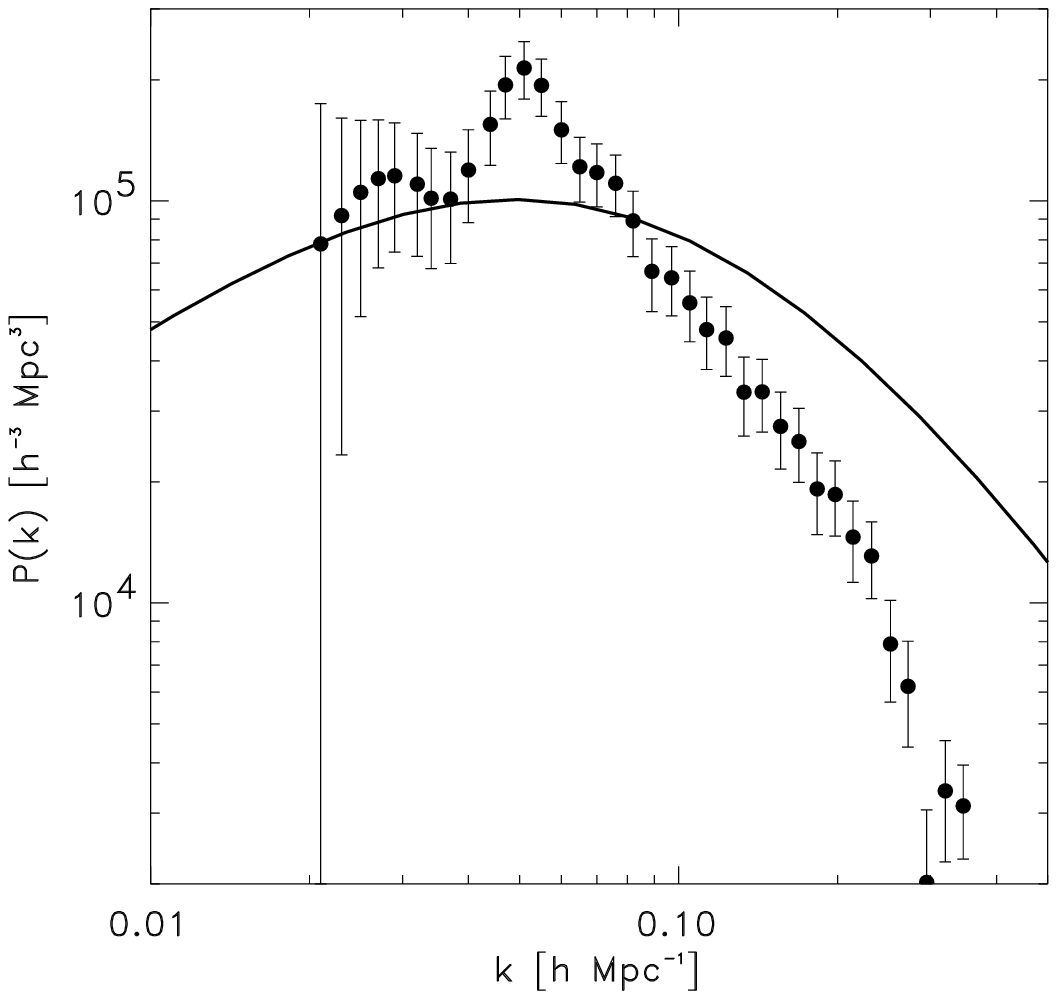}}
\noi
{ FIG.~2.} The power spectrum $P(k)$ for 869 clusters with measured
redshifts is plotted with solid circles. The spectrum is calculated
from the cluster correlation function via the Fourier transform,
both functions are found in redshift space.
Errors were determined from $2\sigma$ errors of the correlation
function found from the scatter of different simulations. 
As for each wavenumber we integrate over the whole interval of the
correlation function individual values of the spectrum are not
independent of each other. To check the indirect method of calculation
of the power spectrum we have used simulated cluster samples having
similar selection effects as real ones. This check was performed for a
wide variety of models with different initial spectra. Our results
show that the true spectrum can be restored over the wavenumber
interval from $k\approx 0.03$~$h$~Mpc$^{-1}$ towards shorter waves
until $k\approx 0.3$~$h$~Mpc$^{-1}$.  The solid line is the standard
CDM ($h =0.5$, $\Omega = 1$) power spectrum enhanced by a bias
factor of $b=3$ over the four year COBE normalisation.
\endinsert

In both methods the main problem is the calculation of the selection
function of clusters of galaxies which corrects for incompleteness
both at low Galactic latitude $b$ and at large distances $r$ from the
observer. The selection function can be represented by linear
functions of $\sin b$ and $r$ (for numerical values of selection
function parameters see refs 12 and 15). Both methods to derive the power
spectrum yield similar results. \dlt{However, parameters of the
correlation function are insensitive to small inaccuracies of the
selection function$^{15}$, thus the indirect method yields more
accurate results for the power spectrum.}

The power spectrum for clusters of galaxies is shown in Fig.~2.
\dlt{This power spectrum is among the first for three-dimensional data 
having measurements on scales well above 100~\Mpc.} On very large
scales the errors are large due to incomplete data. On moderate scales
we see one single well defined peak at a wavenumber $k_0=0.052~h~{\rm
Mpc}^{-1}$.  Errors are small near the peak, and the relative
amplitude and position of the peak are determined \dlt{quite}
accurately. The wavelength of the peak is $\lambda_0=2\pi/k_0=120 \pm
15$~\Mpc. \dlt{Near the peak, there is an excess in the amplitude of
the observed power spectrum over that of the \ed{cold dark matter} (CDM)
 model (see below) by
a factor of 1.4.} Within observational errors our power spectrum on
large scales is compatible with the Harrison-Zeldovich spectrum with
constant power index $n=1$, and on small scales with a spectrum of
constant negative power index $n=-1.8$.

The power spectrum is often used to compare the distribution of matter in the
Universe with theoretical predictions. \dlt{Currently popular structure
formation theories based on the dynamics of a Universe dominated by CDM yield
spectra of which one example is plotted in Figure~2 as a solid line.  The
spectrum is rising on long wavelengths $\lambda$ (small values of the
wavenumber $k=2\pi/\lambda$), and falling \ed{at short} wavelengths (large
values of $k$)}.  The transition between \ed{short-} and long-wavelength
regions in the CDM-spectrum is smooth. The distribution of superclusters in
CDM-models is irregular$^{16}$.

As we see, the relative amplitude of the observed power spectrum above
the standard CDM-type model is not very large. Thus we may ask the
question: \dlt{within the framework of the standard cosmogony, how}
frequently can we expect to \dlt{find} a distribution of clusters
which has a power spectrum similar to \dlt{that observed}? To answer
this question we determined the correlation function and power
spectrum for clusters in rich superclusters \dlt{of} CDM-type
models. In the spectral range of interest the power spectrum of the
standard \dlt{CDM model} is similar to the spectrum of a random
supercluster model$^{14}$. \dlt{We make use of this similarity by
generating 1,000 realizations of the random supercluster model,
applying the selection effects as found in cluster distribution, and
determining the parameters of the cluster correlation function and
power spectrum}.  We measure the mean period and amplitude of the
oscillating correlation function and their respective scatter.  We
also calculate the deviations for individual periods.  This test shows
that \dlt{ combination of parameters} (for instance period vs.
amplitude) close to \dlt{the} observed values \dlt{occurs in}
approximately 1~\% of cases, but the \dlt{simultaneous concurrence of
all parameters with observations} is a much more \dlt{rare}
event. Thus some change in the initial spectrum of matter is
\dlt{necessary in order} to explain the observed correlation function
and power spectrum \dlt{for} clusters of galaxies.

The regularity of the distribution of superclusters is quite striking
and has some similarity with the regularity found in pencil-beam galaxy
surveys$^1$.  Independent evidence for the presence of a preferred
scale in the Universe around \dlt{100~\Mpc} comes from a core-sampling
analysis of the distribution of sheet-like and filamentary 
structures$^{17}$, and from
a power spectrum analysis of \dlt{the} two-dimensional distribution of
galaxies$^{18}$ in the Las Campanas Redshift \dlt{Survey}. Available
data, however, are insufficient to say whether \ed{one-, two-, and our 
three-dimensional} surveys
measure identical physical scales in the Universe, or, if so, whether
differences in numerical values of the scale are due to differences in
methods applied in determination or to a real cosmic variance in the
scale.

According to our present understanding of the Universe the observed
distribution of luminous matter is strongly correlated with the
density perturbations of some dark nonbaryonic matter component(s) at
the moment of recombination. These perturbations are due to processes
in the very early Universe so that their power spectrum at
recombination depends both on these processes and the further
evolution of the Universe, in particular the transition from radiation
to matter dominated expansion. We have shown that the luminous matter
in galaxy clusters is much more regularly distributed than
expected. Thus we end up with the conclusion that our present
understanding of structure formation on very large scales needs
revision.

\bsn
ACKNOWLEDGEMENTS. We thank P.J.E. Peebles for discussion and valuable
suggestions. This study was supported by grants of the International
Science Foundation and Estonian Science Foundation; visits of J.E. and
A.S. to Potsdam were supported by the Deutsche Forschungsgemeinschaft.
A.S. was partially supported by the Russian Research Project
``Cosmomicrophysics''.

\bsn
CORRESPONDENCE should be addressed to J.E. (e-mail: einasto@aai.ee).
\bsn
\hrule
\bsn
Received 10 October; accepted 19 November 1996.

\bsn
\ref  1. Broadhurst, T.J., Ellis, R.S., Koo, D.C. and Szalay, A.S.,
{\it Nature}, {\bf 343},  726 -- 728 (1990).
\ref  2. Kaiser, N., Peacock, J.A., {\it Astrophys. J.}, {\bf 379}, 482
-- 506 (1991).
\ref 3. Kopylov, A. I., Kuznetsov D. Y., Fetisova T. S., and
Shvarzman V. F. , in {\it Large Scale Structure of the Universe},
eds. J. Audouze, M.--C. Pelletan, A. Szalay (Kluwer), p. 129 -- 137 (1988).
\ref 4. Mo H.J., Deng Z.G., Xia X.Y., Schiller P., and B\"orner G.
{\it   Astron. Astrophys.},  {\bf 257}, 1 -- 10 (1992).
\ref 5. Fetisova, T. S., Kuznetsov, D. Y., Lipovetsky, V. A.,
Starobinsky, A. A., and Olowin, R. P. , {\it Astron. Lett.},
{\bf 19}, 198 -- 202 (1993).
\ref 6. Einasto, J., Gramann, M. , {\it Astrophys. J.}, {\bf 407}, 443 
-- 447 (1993).
\ref 7. Andernach, H., Tago, E., and Stengler-Larrea, E., {\it Astroph. 
Lett. \& Comm.} {\bf 31}, 27 -- 30 (1995).
\ref 8. Abell, G., {\it Astrophys. J. Suppl.} {\bf 3}, 211 -- 288 (1958).
\ref 9. Abell, G., Corwin, H., Olowin, R., {\it Astrophys. J. Suppl.}
{\bf 70}, 1 --  138  (1989).
\ref 10. Peacock, J., and West, M.J., {\it Mon. Not. R. astr. Soc.}, 
{\bf 259}, 494 -- 504  (1992)
\ref 11. Einasto, M., Einasto, J., Tago, E., Dalton, G., and Andernach, H.,
{\it Mon. Not. R. Astr. Soc.}, {\bf 269}, 301 --  322 (1994).
\ref 12. Einasto, M., Tago, E., Jaaniste, J., Einasto, J., \&
Andernach, H.,   {\it Astron. Astrophys.} (in the press, SISSA preprint 
astro-ph/9610088) (1996).
\ref 13. Tully, R. B., Scaramella, R., Vettolani, G., and Zamorani, G.
{\it Astroph. J.}, {\bf 388}, 9 -- 16 (1992).
\ref 14. Einasto, J., Einasto, M., Frisch, P., Gottl\"ober, S., M\"uller,
V., Saar, V., Starobinsky, A.A., Tucker, D., 
{\it Mon. Not. R. astr. Soc.} (submitted) (1997).
\ref 15. Einasto, J., Einasto, M., Frisch, P., Gottl\"ober, S., M\"uller,
V., Saar, V., Starobinsky, A.A., Tago, E., Tucker, D., Andernach, H.,
{\it Mon. Not. R. astr. Soc.} (submitted) (1997).
\ref 16. Frisch, P., Einasto, J., Einasto, M., Freudling, W., Fricke, K.J.,
Gramann, M., Saar, V., Toomet, O., {\it Astron. Astrophys.} {\bf 296},
611 -- 627 (1995).
\ref 17. Doroshkevich, A. G, Tucker, D. L., Oemler, A., Kirshner, R. P., Lin,
H., Shectman, S. A., Landy, S.D., \& Fong, R., {\it Mon. Not. R. astr. Soc.}
(submitted) (1996).
\ref 18. Landy, S.D., Shectman, S.A., Lin, H., Kirshner, R.P., Oemler,
A.A., Tucker, D., {\it Astrophys. J.}, {\bf 456}, L1 -- L4 (1996).

\bye